\documentclass[showpacs,eqsecnum,aps,twocolumn,floatfix]{revtex4}
\usepackage{graphicx,subfigure,amsmath,amssymb}
\usepackage{epsfig}
\usepackage{multirow}
\usepackage{rotating}

\begin{document}

\title{Proton emission with a screened electrostatic barrier}

\author{R. Budaca$^{1,2}$ and A. I. Budaca$^{1}$}

\affiliation{$^{1)}$"Horia Hulubei" National Institute for Physics and Nuclear Engineering, Str. Reactorului 30, RO-077125, POB-MG6, M\v{a}gurele-Bucharest, Romania}
\affiliation{$^{2)}$Academy of Romanian Scientists, 54 Splaiul Independenei, RO-050094, Bucharest, Romania}

\begin{abstract}
Half-lives of proton emission for $Z\geq51$ nuclei are calculated within a simple analytical model based on the WKB approximation for the barrier penetration probability which includes the centrifugal and overlapping effects besides the electrostatic repulsion. The model has a single free parameter associated to a Hulthen potential which emulates a Coulomb electrostatic interaction only at short distance. The agreement with experimental data is very good for most of the considered nuclei. Theoretical predictions are made for few cases with uncertain emitting state configuration or incomplete decay information. The model's assignment of the proton orbital momentum is in agreement with the differentiation of the experimental data by orbital momentum values realized with a newly introduced correlation formula.
\end{abstract}

\maketitle

\section{Introduction}

Proton radioactivity is understood as the disintegration of nuclei by the emission of a proton and is specific to proton rich odd-$Z$ nuclei. Since its first observation in an isomeric state of $^{53}$Co by Jackson {\it et al.} \cite{Jackson}, and immediate confirmation by Cerny {\it et al.} \cite{Cerny}, the proton emission became an invaluable source of detailed nuclear structure information for nuclides far from the $\beta$-stability line. The limit at which nuclei become unbound to the emission of a proton from their ground states defines the so-called proton drip-line which is a fundamental guideline for nucleosynthesis. Indeed, the proton drip-line put some narrow constraints for the synthesis  of proton rich nuclei in explosive astrophysical scenarios such as X-ray bursts \cite{Schatz} and neutrino driven winds \cite{Wanajo}, where the inverse process of proton radioactivity, called rapid proton capture ($rp$), plays an essential role \cite{Wallace}. The study of proton radioactivity is therefore essential for mapping the proton drip-line \cite{Woods}, especially since most of the observed proton emitters are found in the range $Z\geq51$ \cite{Blank,Pfu} were the proton drip-line is not well defined.

For proton emission to occur, the condition of negative proton separation energy is not enough because the odd proton must penetrate a potential barrier corresponding to an electrostatic interaction as well as a centrifugal contribution. The later have a more important role in comparison to $\alpha$ and cluster decays, due to the much smaller mass of the proton. Such an interplay between the electrostatic and centrifugal barriers cause $Z\geq51$ nuclei beyond the proton drip-line to survive long enough to be detected with half-lives ranging from $10^{-6}$ s to a few seconds. In contradistinction, low $Z$ proton rich nuclei cannot be detected directly, being instead registered just as short-lived resonances.

The theoretical description of this rare phenomenon received much attention in the recent decade through the natural extension of models well established for $\alpha$ and cluster decays. These traditionally include phenomenological \cite{Dong,Raj,Guo,Qian}, microscopic \cite{Fer,Zhao,Lim} and semi-classical formalisms \cite{Medeiros,Qi,Ni,Zdeb}. The involved phenomenological and microscopic approaches include many fitted parameters and gross approximations or rely on the spectroscopic information regarding the single-particle configuration of the decaying state, which is unfortunately lacking for the most of the proton emitting nuclei. In view of these shortcomings and due to increased number of measured proton emissions the simple semi-classical methods based on the WKB approximation provide not only a reliable quantitative description for the proton emission using a minimal number of parameters, but also a clear physical meaning for the decay ingredients. In this study we pursue the same reasoning and apply the WKB analysis to a potential barrier completely determined by a single parameter associated to the range of the electrostatic interaction. By considering a Hulthen potential \cite{Hulthen1,Hulthen2} for the electrostatic barrier we can account for additional short range effects, such as proximity nuclear interaction and charge diffuseness. Indeed, matching the outer turning point of the Hulthen and Coulomb potentials amounts to an increase in the usual Coulomb barrier at short distance. From the successful reproduction of experimental data with such a simple approach one will be able to ascertain the validity of specific approximations and the relative importance of the ignored structural features and secondary effects. Not least important for the decay studies are the empirical decay laws. A new such correlation will be also introduced here as a supporting test for the model predictions.

\renewcommand{\theequation}{2.\arabic{equation}}
\setcounter{equation}{0}
\section{The theoretical framework}

From the quantum mechanical point of view, proton emission can be modeled as the tunneling of a valence proton through a potential barrier. Here one separates the barrier into inner and outer regions in terms of the nuclear radius $r$. The first region is very narrow and contains information about the transition of the proton from within the compound parent nucleus to the touching configuration. It is defined by the interval between the radius of the parent nucleus $R_{0}$ and the distance of the touching configuration $R_{t}=R_{1}+r_{p}$, where $R_{1}$ and $r_{p}$ are the radii of the daughter nucleus and of the proton, respectively. The proton radius is considered to be 0.84 fm, while the hard nuclear radii are defined by
\begin{equation}
R_{i}=1.28A_{i}^{1/3}-0.76+0.8A_{i}^{-1/3},\,\,i=0,1,
\end{equation}
where $A_{0}$ and $A_{1}$ are the mass numbers corresponding to the parent and respectively daughter nuclei. The potential for this inner preformation part is parametrized as \cite{Raj,Bala}
\begin{equation}
V_{in}(r)=a_{1}r+a_{2}r^{2}.
\end{equation}
The constants $a_{1}$ and $a_{2}$ are fixed by requiring $V_{in}(R_{0})=Q_{p}$ and matching the inner and outer potentials at $R_{t}$. The introduction of this inner barrier serves as a phenomenological counterpart for the spectroscopic factor which defines the proton preformation probability in terms of single-particle level occupancies. In what concerns the outer barrier, it is defined as a superposition of a centrifugal energy term
\begin{equation}
V_{l}(r)=\frac{\hbar^{2}l(l+1)}{2\mu r^{2}}
\end{equation}
and a repulsive electrostatic potential. $\mu=mA_{1}/(A_{1}+1)$ is the reduced mass of the decaying nuclear system with $m$ being the nucleon mass. The orbital momentum $l$ of the emitted proton must satisfy the angular momentum and parity conservation laws concerning the initial and final nuclear states. The electrostatic potential is by default of the Coulomb type $V_{C}(r)=Z_{1}e^{2}/r$, where $Z_{1}$ is the charge number of the daughter nucleus. However, in this study we will employ a Hulthen \cite{Hulthen1,Hulthen2} type potential,
\begin{equation}
V_{H}(r)=\frac{a e^{2}Z_{1}}{e^{ar}-1},
\end{equation}
which is actually a generalization of the Coulomb potential with a screening effect included by means of the parameter $a$. Contrary to the Coulomb potential, Hulthen potential is of a short range, behaving as a Coulomb potential at short distance and dropping exponentially at large distance. The Hulthen potential is very important in atomic, molecular and solid state physics where the bound or free electrons play an important role in the configuration of the electrostatic field. In the present case however, one cannot speak of an electronic screening because we deal with bare nuclei, and moreover the mass (energy) range of the proton is beyond that of electrons. Nevertheless, a Hulthen potential allows to adjust the Coulomb potential by means of its convergence range $a$, which can be considered to account for the finite size nuclear effects in a gross manner. The deviations from the electrostatic approximation, {\it i.e.} the superposition of the involved charges, movement of the proton which generates a magnetic field and the inhomogeneous charge distribution of the nucleus, also bespeak for a reconsideration of the Coulomb potential. Moreover, the general theory of scattering is immediately applicable to the case of the Hulthen potential \cite{Hulthen2}, which is not the case of the Coulomb potential because it decreases too slowly to infinity.

The proton decay half-live is generically defined as
\begin{equation}
T_{1/2}=\frac{ln2}{\nu_{p}P},
\end{equation}
where $P$ is the probability of the proton to penetrate a phenomenological potential barrier, while
\begin{equation}
\nu_{p}=\frac{1}{2R_{0}}\sqrt{\frac{2E_{p}}{\mu}}
\end{equation}
is the proton assault frequency on the barrier. $E_{p}$ is the measured kinetic energy of the proton related to the total decay energy $Q_{p}$ shared between the proton and the recoiling atom by \cite{Pfu}:
\begin{equation}
Q_{p}=\frac{m_{p}+M(N,Z-1)+m_{e}}{M(N,Z-1)+m_{e}}E_{p},
\label{Ep}
\end{equation}
where $m_{p}=1.007$ a.u. and $m_{e}=5.486\cdot10^{-4}$ a.u. are the proton and electron masses. The barrier penetrability is calculated by means of the WKB approximation:
\begin{equation}
P=Exp\left\{-\frac{2}{\hbar}\int_{R_{in}}^{R_{out}}\sqrt{2\mu\left[V(r)-Q_{p}\right]}dr\right\},
\end{equation}
where $R_{in}=R_{0}$, while $R_{out}$ is the second turning point defined by $V_{out}(R_{out})=Q_{p}$. The exponent $G=-\log{P}$ is the well known Gamow factor. As the total potential is separated in two regions
\begin{equation}
V(r)=\left\{\begin{array}{l}
V_{in},\,\,r<R_{t}\\
V_{out}=V_{H}(r)+V_{l}(r),\,\,r>R_{t},
\end{array}\right.
\end{equation}
the barrier penetrability can be factorized as $P=P_{in}P_{out}$, with associated Gamow factors
\begin{eqnarray}
G_{in}&=&\frac{2}{\hbar}\int_{R_{0}}^{R_{t}}\sqrt{2\mu\left[V_{in}(r)-Q_{p}\right]}dr,\\
G_{out}&=&\frac{2}{\hbar}\int_{R_{t}}^{R_{out}}\sqrt{2\mu\left[V_{out}(r)-Q_{p}\right]}dr.
\end{eqnarray}
The first factor have a simple analytic form
\begin{eqnarray}
&&G_{in}=\left(\frac{a_{1}}{4a_{2}}+\frac{R_{t}}{2}\right)\sqrt{a_{1}R_{t}+ a_{2}R_{t}^{2}-Q_{p}}-\nonumber\\
&&\frac{1}{\sqrt{a_{2}}}\left(\frac{a_{1}^{2}}{8a_{2}}+\frac{Q_{p}}{2}\right)\\
&&\times \log{\left[\frac{2\sqrt{a_{2}\left(a_{1}R_{t}+ a_{2}R_{t}^{2}-Q_{p}\right)}+a_{1}+2a_{2}R_{t}}{a_{1}+2a_{2}R_{0}}\right]}\nonumber,
\end{eqnarray}
where the parameters of $V_{in}(r)$ are expressed as
\begin{equation}
a_{k}=(-)^{k}\frac{Q_{p}R_{t}^{3-k}-V_{out}(R_{t})R_{0}^{3-k}}{R_{t}R_{0}(R_{0}-R_{t})},\,\,k=1,2.
\end{equation}

To calculate the Gamow factor for the outer region of the barrier, one must first amend the centrifugal term by the Langer correction, {\it i.e.} to replace $l(l+1)$ with $(l+1/2)^{2}$ \cite{Langer}. This modification necessarily arises in the WKB approximation when the spherical symmetry of the system is assumed. The spherical symmetry of the proton emission phenomenon is essential in order for the angular momentum associated to the proton to be a good quantum number. The consequences of Langer transform are not negligible, being especially important for the $l\neq0$ case \cite{Lim}. $G_{out}$ can be analytically determined by rewriting the centrifugal term as \cite{Filho}:
\begin{equation}
\frac{1}{r^{2}}\approx\frac{a^{2}}{\left(e^{ar}-1\right)^2}.
\end{equation}
The above approximation retains the functional form of the potential and is very good for small values of $a$ such that its radius of validity is much extended in comparison to the region of superposition between a Coulomb potential and a Hulthen potential with the same screening parameter. With this approximation in place, the barrier exit radius can be expressed as follows:
\begin{equation}
R_{out}=\frac{1}{a}\log{\left[\frac{2V_{1}}{\sqrt{V_{0}^{2}+4V_{1}Q_{p}}-V_{0}}+1\right]},
\end{equation}
where
\begin{equation}
V_{0}=ae^{2}Z_{1},\,\,V_{1}=\frac{a^{2}\hbar^{2}\left(l+\frac{1}{2}\right)^{2}}{2\mu}.
\end{equation}
The Gamow factor for the outer barrier region, in the same approximation is analytically given by:
\begin{equation}
G_{out}(r)=\frac{1}{a}\left[I_{1}(r)+I_{2}(r)\right]\Big|_{R_{t}}^{R_{out}}.
\end{equation}
The two terms have similar expressions:
\begin{eqnarray}
&&I_{1}(r)=-\sqrt{V_{1}x^{2}+V_{0}x-Q_{p}}+\\
&&\sqrt{Q_{p}}\arcsin{\left[\frac{xV_{0}-2Q_{p}}{x\sqrt{4Q_{p}V_{1}+V_{0}^{2}}}\right]}-\nonumber\\
&&\frac{V_{0}}{2\sqrt{V_{1}}}\log{\left[2\sqrt{V_{1}\left(V_{1}x^{2}+V_{0}x-Q_{p}\right)}+V_{0}+2V_{1}x\right]}\nonumber,
\end{eqnarray}
with $x=(e^{ar}-1)^{-1}$, and
\begin{eqnarray}
&&I_{2}(r)=\sqrt{V_{1}y^{2}+U_{0}y-U_{1}}-\nonumber\\
&&\sqrt{U_{1}}\arctan{\left[\frac{yU_{0}-2U_{1}}{2\sqrt{U_{1}\left(V_{1}y^{2}+U_{0}y-U_{1}\right)}}\right]}+\nonumber\\
&&\frac{U_{0}}{2\sqrt{V_{1}}}\log{\left[2\sqrt{V_{1}\left(V_{1}y^{2}+U_{0}y-U_{1}\right)}+U_{0}+2V_{1}y\right]}\nonumber\\
\end{eqnarray}
where $y=1+(e^{ar}-1)^{-1}$ and the following notations were used
$U_{0}=V_{0}-2V_{1},\,\,U_{1}=Q_{p}+V_{0}-V_{1}$.

\renewcommand{\theequation}{3.\arabic{equation}}
\setcounter{equation}{0}
\section{Numerical results}

\begin{figure}[t!]
\begin{center}
\includegraphics[clip,trim = 0mm 0mm 0mm 0mm,width=0.47\textwidth]{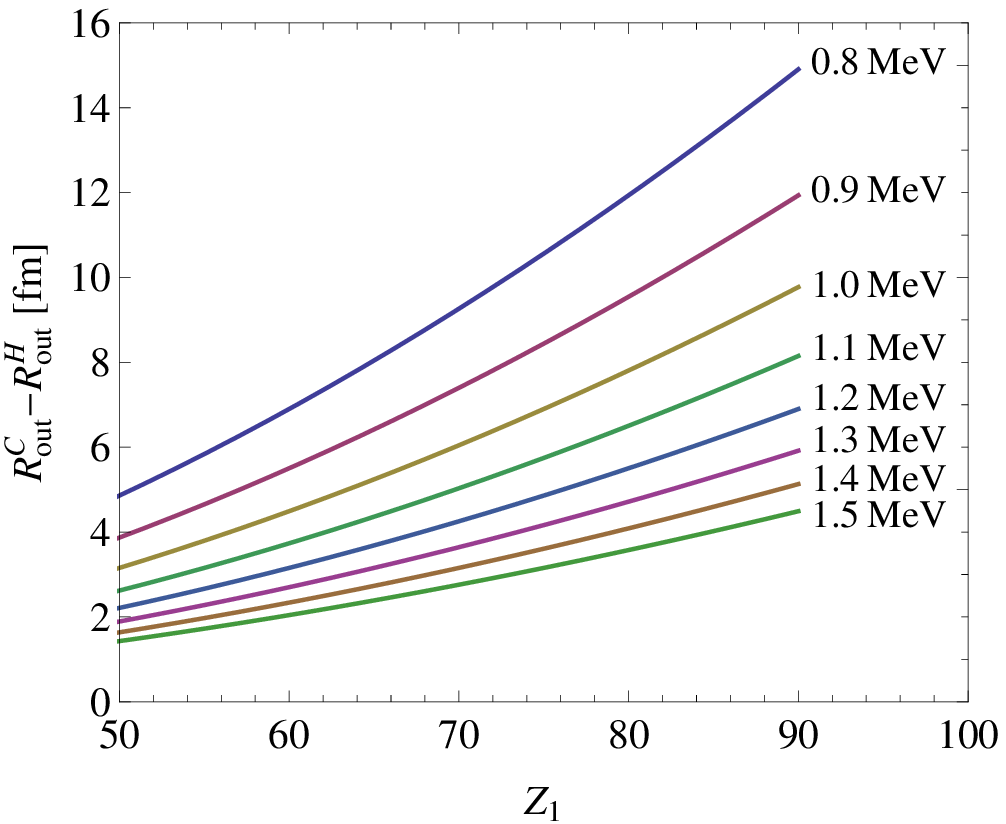}
\caption{The difference between the turning points associated to Coulomb potential and Hulthen potential with $a=1.299\cdot10^{-3}$ fm$^{-1}$, plotted as a function of the charge number of the daughter nucleus for different values of decay energy $Q_{p}$. The turning point radii are defined by $V_{i}(R_{i})=Q_{p},(i=C,H)$.}
\label{dR}
\end{center}
\end{figure}

The proposed model has a single free parameter, the screening parameter $a$ which is adjusted to fit the experimental data. The experimental data used in the fitting procedure corresponds to 41 observed and measured ground state and isomeric proton emissions from nuclei with $Z>50$ for which all the needed information such as decay energy, angular momentum, branching ratio and half-lives are known and assigned without major uncertainties. The value of $a$ is then found by minimizing the quantity
\begin{equation}
\sigma=\sqrt{\frac{1}{41}\sum_{i=1}^{41}\left[\log\left(\frac{T_{th}^{i}}{T_{exp}^{i}}\right)\right]^{2}},
\end{equation}
which is just the standard deviation. Due to the analytic structure of the formalism the fitting procedure is straightforward and provides $a=1.299\cdot10^{-3}$ fm$^{-1}$ corresponding to $\sigma=0.418$. As was expected, the value of $a$ is quite small. This suggests that the electrostatic hypothesis of the usually employed Coulomb potential is a fairly good approximation. Nevertheless, the effect of non zero screening in the description of the proton emission phenomenon is sizable as can be seen in Fig. \ref{dR}, where one plotted the difference between the outer turning point radii corresponding to pure Coulomb and Hulthen barriers, {\it i.e.} without the centrifugal contribution. In case of Coulomb barrier, this radius takes values between 70 and 115 fm for the considered nuclei. The screening of the electrostatic repulsion shortens this radius by several percents. The squeezing of the barrier is obviously more pronounced for lower reaction energies, where the dependence on the charge number of the final nucleus is also enhanced.

\setlength{\tabcolsep}{9.5pt}
\begin{table*}[t!]
\begin{center}
\caption{Decay properties of measured proton emitters with $Z\geq51$: proton emitting nucleus, orbital momentum transferred by the proton, decay energy $Q_{p}$ mostly collected from \cite{Wang} or extracted from more recently measured kinetic energies $E_{p}$, partial proton emission half-live and its decimal logarithm, and the origin of the data concerning orbital momentum and half-live. ($\ast$) denotes an isomeric state for a nucleus. The theoretical half-lives and their logarithmic representations obtained with the present formalism (th1) as well as with UDLP (\ref{udlp}) (th2) are also listed for comparison.}
\label{tab:1}
{\footnotesize
\begin{tabular}{lccccccrrr}
\hline
Nucleus & $l$ & $Q_{p}$ & Ref. & \multicolumn{3}{c}{$T_{1/2}$} & \multicolumn{3}{c}{$\log_{10}{\left[T_{1/2}(s)\right]}$}\\
\cline{5-7}\cline{8-10}
        &     & [MeV]   &      & exp       & th1 & th2        &  exp & th1 & th2\\
\hline
$^{105}_{\,\,\,51}$Sb      & 2 &0.4830 & \cite{105Sb}       &122(11) s            &59 s        &76 s        & 2.086& 1.768& 1.881\\
$^{109}_{\,\,\,53}$I       & 2 &0.8195 & \cite{Petri,109I}  &92.8(8) $\mu$s       &85.3 $\mu$s &184.8 $\mu$s&-4.032&-4.069&-3.733\\
$^{112}_{\,\,\,55}$Cs      & 2 &0.8160 & \cite{Audi}        &490(35) $\mu$s       &467 $\mu$s  &856 $\mu$s  &-3.310&-3.330&-3.068\\
$^{113}_{\,\,\,55}$Cs      & 2 &0.9735 & \cite{113Cs}       &17.7(4) $\mu$s       &4.4 $\mu$s  &11.0 $\mu$s &-4.752&-5.360&-4.957\\
$^{130}_{\,\,\,63}$Eu      & 2 &1.0280 & \cite{Dav}         &0.90$^{+49}_{-29}$ ms&0.26 ms     &0.39 ms     &-3.046&-3.585&-3.404\\
$^{131}_{\,\,\,63}$Eu      & 2 &0.9470 & \cite{Audi}        &20.0(29) ms          &2.7 ms      &3.5 ms      &-1.699&-2.577&-2.461\\
$^{135}_{\,\,\,65}$Tb      & 3 &1.1880 & \cite{Audi}        &1.01(28) ms          &0.12 ms     &0.17 ms     &-2.996&-3.906&-3.770\\
$^{140}_{\,\,\,67}$Ho      & 3 &1.0940 & \cite{140Ho}       &6(3) ms              &4.4 ms      &4.4 ms      &-2.222&-2.360&-2.359\\
$^{141}_{\,\,\,67}$Ho      & 3 &1.1770 & \cite{141Ho}       &4.1(1) ms            &0.5 ms      &0.6 ms      &-2.387&-3.280&-3.214\\
$^{141}_{\,\,\,67}$Ho$^{*}$& 0 &1.2430 & \cite{141Ho}       &7.3(3) $\mu$s        &2.2 $\mu$s  &5.1 $\mu$s  &-5.137&-5.649&-5.291\\
$^{145}_{\,\,\,69}$Tm      & 5 &1.7360 & \cite{Audi}        &3.17(20) $\mu$s      &10.37 $\mu$s&20.04 $\mu$s&-5.499&-4.984&-4.698\\
$^{146}_{\,\,\,69}$Tm      & 0 &0.8960 & \cite{Audi,Rob}    &155(20) ms           &155 ms      &170 ms      &-0.810&-0.810&-0.771\\
$^{146}_{\,\,\,69}$Tm$^{*}$& 5 &1.2000 & \cite{Audi,Rob}    &75(7) ms             &205 ms      &154 ms      &-1.125&-0.687&-0.812\\
$^{147}_{\,\,\,69}$Tm      & 5 &1.0590 & \cite{Audi}        &3.87(130) s          &8.17 s      &4.64 ms     & 0.587& 0.912& 0.667\\
$^{147}_{\,\,\,69}$Tm$^{*}$& 2 &1.1210 & \cite{Audi}        &0.36(4) ms           &1.08 ms     &1.27 ms     &-3.444&-2.966&-2.897\\
$^{150}_{\,\,\,71}$Lu      & 5 &1.2696 & \cite{150Lu}       &63(5) ms             &125 ms      &92 ms       &-1.197&-0.904&-1.037\\
$^{150}_{\,\,\,71}$Lu$^{*}$& 2 &1.2916 & \cite{150Lu}       &39$^{+8}_{-6}$ $\mu$s&64 $\mu$s   &87 $\mu$s   &-4.409&-4.191&-4.059\\
$^{151}_{\,\,\,71}$Lu      & 5 &1.2407 & \cite{151Lu,151Lu2}&122(2) ms            &237 ms      &165 ms      &-0.914&-0.626&-0.782\\
$^{151}_{\,\,\,71}$Lu$^{*}$& 2 &1.2940 & \cite{151Lu3}       &17(1) $\mu$s         &60 $\mu$s   &81 $\mu$s   &-4.770&-4.224&-4.090\\
$^{155}_{\,\,\,73}$Ta      & 5 &1.4530 & \cite{Audi}        &3.2(13) ms           &7.5 ms      &6.6 ms      &-2.495&-2.128&-2.180\\
$^{156}_{\,\,\,73}$Ta      & 2 &1.0200 & \cite{Audi}        &149(8) ms            &295 ms      &222 ms      &-0.826&-0.531&-0.654\\
$^{156}_{\,\,\,73}$Ta$^{*}$& 5 &1.1140 & \cite{Audi}        &8.57(207) s          &19.66 s     &9.32 s      & 0.933& 1.294& 0.969\\
$^{157}_{\,\,\,73}$Ta      & 0 &0.9350 & \cite{157Ta}       &0.30(16) s           &0.75 s      &0.67 s      &-0.527&-0.125&-0.174\\
$^{159}_{\,\,\,75}$Re$^{*}$& 5 &1.8160 & \cite{159Re}       &21.6(44) $\mu$s      &44.9 $\mu$s &61.6 $\mu$s &-4.665&-4.348&-4.211\\
$^{160}_{\,\,\,75}$Re      & 2 &1.2670 & \cite{160Re}       &0.90$^{+17}_{-10}$ ms&1.25 ms     &1.24 ms     &-3.045&-2.905&-2.908\\
$^{161}_{\,\,\,75}$Re      & 0 &1.1970 & \cite{Audi}        &440(1) $\mu$s        &1.08 ms     &1.28 ms     &-3.357&-2.968&-2.891\\
$^{161}_{\,\,\,75}$Re$^{*}$& 5 &1.3207 & \cite{Audi}        &210.0(10) ms         &327.5 ms    &198.6 ms    &-0.678&-0.485&-0.702\\
$^{164}_{\,\,\,77}$Ir$^{*}$& 5 &1.8253 & \cite{164Ir}       &73(11) $\mu$s        &93 $\mu$s   &111 $\mu$s  &-4.137&-4.034&-3.953\\
$^{165}_{\,\,\,77}$Ir$^{*}$& 5 &1.7200 & \cite{164Ir}       &0.386(46) ms         &0.452 ms    &0.465 ms    &-3.413&-3.345&-3.333\\
$^{166}_{\,\,\,77}$Ir      & 2 &1.1520 & \cite{Audi}        &150.0(716) ms        &82.7 ms     &59.9 ms     &-0.824&-1.083&-1.223\\
$^{166}_{\,\,\,77}$Ir$^{*}$& 5 &1.3240 & \cite{Audi}        &0.84(28) s           &0.89 s      &0.47 s      &-0.076&-0.050&-0.326\\
$^{167}_{\,\,\,77}$Ir      & 0 &1.0700 & \cite{Audi}        &74.6(2.9) ms         &145.4 ms    &124.0 ms    &-1.128&-0.838&-0.907\\
$^{167}_{\,\,\,77}$Ir$^{*}$& 5 &1.2480 & \cite{167Ir}       &6.9(13) s            &5.4 s       &2.5 s       & 0.836& 0.735& 0.401\\
$^{170}_{\,\,\,79}$Au      & 2 &1.4720 & \cite{Audi}        &326(67) $\mu$s       &140 $\mu$s  &146 $\mu$s  &-3.487&-3.854&-3.835\\
$^{170}_{\,\,\,79}$Au$^{*}$& 5 &1.7520 & \cite{Audi}        &1.07(13) ms          &0.67 ms     &0.62 ms     &-2.971&-3.177&-3.206\\
$^{171}_{\,\,\,79}$Au      & 0 &1.4480 & \cite{Audi}        &22.3(24) $\mu$s      &37.4 $\mu$s &49.8 $\mu$s &-4.652&-4.427&-4.303\\
$^{171}_{\,\,\,79}$Au$^{*}$& 5 &1.7030 & \cite{Audi}        &2.6(2) ms            &1.4 ms      &1.2 ms      &-2.587&-2.843&-2.903\\
$^{176}_{\,\,\,81}$Tl      & 0 &1.2650 & \cite{Audi}        &6.2(23) ms           &8.3 ms      &7.4 ms      &-2.208&-2.079&-2.130\\
$^{177}_{\,\,\,81}$Tl      & 0 &1.1600 & \cite{Audi}        &67(37) ms            &143 ms      &109 ms      &-1.176&-0.844&-0.964\\
$^{177}_{\,\,\,81}$Tl$^{*}$& 5 &1.9670 & \cite{Audi}        &353(130) $\mu$s      &61 $\mu$s   &68 $\mu$s   &-3.452&-4.214&-4.166\\
$^{185}_{\,\,\,83}$Bi$^{*}$& 0 &1.6070 & \cite{185Bi}       &64(5) $\mu$s         &14 $\mu$s   &18 $\mu$s   &-4.191&-4.848&-4.742\\
\hline
\end{tabular}}
\end{center}
\end{table*}

\begin{figure}[t!]
\begin{center}
\includegraphics[clip,trim = 38mm 0mm 98mm 0mm,width=0.47\textwidth]{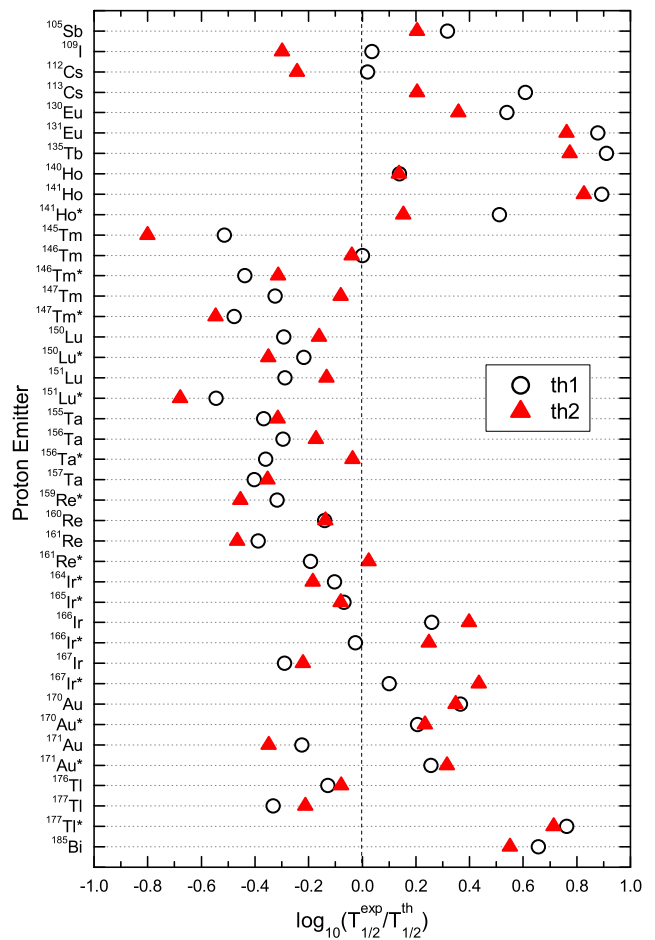}
\caption{The deviations between calculated and measured decimal logarithms of proton emission half-lives for the considered nuclei. Open circles refer to results obtained with the present model, while red triangles denote the UDLP deviations.}
\label{dev}
\end{center}
\end{figure}

The comparison of the theoretical predictions with the experimental data is provided in Table \ref{tab:1}. The resulted rms value is comparable to other approaches which however uses a greater number of adjustable parameters. For example, we fitted the same data with the simple universal decay law for proton emission (UDLP) \cite{Qi}:
\begin{equation}
\log_{10}{T_{1/2}}=A\chi'+B\rho'+C+D\frac{l(l+1)}{\rho'},
\label{udlp}
\end{equation}
where $A,B,C$ and $D$ are free parameters, while the variables are defined as:
\begin{equation}
\chi'=Z_{1}\sqrt{\frac{A_{1}}{(A_{1}+1)Q_{p}}},\,\,\rho'=\sqrt{\frac{A_{1}Z_{1}(A_{1}^{1/3}+1)}{A_{1}+1}}.
\end{equation}
Formula (\ref{udlp}) is based on the simple premise of the quantum tunneling through a centrifugal and Coulomb barriers. Such that its comparison to the present approach would show how adequate are the new ingredients, that is the screening of the electrostatic interaction and the modeling of the pre-scission stage. Thus, it is found that the four parameter fit with UDLP gives an rms value of $\sigma=0.395$, which is barely better than the present model fit with a single parameter. The four parameters extracted from the UDLP fit are $A=0.374$, $B=-0.472$, $C=-17.828$, and $D=2.463$. These values are consistent with previous results \cite{Qi}. The UDLP predictions are also listed in Table \ref{tab:1}, where one can see that although the overall quality of the two fits is similar there are nuclei where the two theoretical predictions significantly diverge. In order to have a more insightful opinion on the relative success of the two approaches, one plotted the corresponding deviations of the $\log_{10}{T_{1/2}}$ between theory and experiment in Fig. \ref{dev}. The major divergence of theoretical results is found for $^{109}$I, $^{145}$Tm, $^{166}$Ir$^{*}$, and $^{167}$Ir$^{*}$ where the present model have an ascendant, and for $^{113}$Cs, $^{141}$Ho$^{*}$, $^{147}$Tm, and $^{156}$Ta$^{*}$ with a better reproduction of data for UDLP. Fig. \ref{dev} also distinguishes two regions: up to $^{159}$Re$^{*}$ where the UDLP formula is predominantly better, and starting from the same nucleus where the present approach becomes more successful in reproducing the data.

The biggest discrepancies within the present calculation are found for $^{131}$Eu, $^{135}$Tb, and $^{141}$Ho nuclei where the experimental $\log[T_{1/2}(s)]$ values are overestimated by the theoretical results with approximately 0.9. The poor reproduction of experimental half-lives for these nuclei is unalterable in other theoretical formulations \cite{Raj,Qian,Medeiros,Qi,Zdeb}, including UDLP. The origin of this inconsistency could be ascribed to the transitional character of these nuclei, which mark the end of the lower $Z$ sequence of strongly prolate emitters \cite{Qi,Delion2,Delion1}. Moreover, the quad\-ru\-pole deformation of the discussed nuclei are among the highest.

\begin{table*}[t!]
\begin{center}
\caption{Same as in Table \ref{tab:1}. For $^{117}$La, $^{121}$Pr, and $^{172}$Au one listed the most probable proton orbital momenta with present (th1) and UDLP (th2) theoretical predictions for each entry.}
\label{tab:2}
{\footnotesize
\begin{tabular}{lccccccrrr}
\hline
Nucleus & $l$ & $Q_{p}$ & Ref. & \multicolumn{3}{c}{$T_{1/2}$} & \multicolumn{3}{c}{$\log_{10}{\left[T_{1/2}(s)\right]}$}\\
\cline{5-7}\cline{8-10}
        &     & [MeV]   &      & exp       & th1         & th2        &  exp & th1& th2\\
\hline
$^{117}_{\,\,\,57}$La      & 2 &0.823  & \cite{117La}   &25.0(28) ms        &1.7 ms     &2.7 ms     &-1.602&-2.762&-2.564\\
                           & 3 &       & \cite{117La2}  &                   &14.9 ms    &18.0 ms    &      &-1.826&-1.746\\
$^{121}_{\,\,\,59}$Pr      & 2 &0.900  & \cite{121Pr}   &10$^{+6}_{-3}$ ms  &0.64 ms    &1.02 ms    &-2.000&-3.192&-2.992\\
                           & 3 &       & \cite{121Pr2}  &                   &5.3 ms     &6.4 ms     &      &-2.277&-2.192\\
$^{144}_{\,\,\,69}$Tm      & 5 &1.712  & \cite{Audi}    &$\geq$2.3(9) $\mu$s&15.3 $\mu$s&28.1 $\mu$s&-5.638&-4.817&-4.551\\
$^{172}_{\,\,\,79}$Au      & 0 &0.810  & \cite{Audi}    &$>$1.4(2) s        &3.7 h      &1.9 h      & 0.146& 4.121& 3.832\\
                           & 2 &       & \cite{172Au}   &                   &22.3 h     &8.5 h      &      & 4.905& 4.487\\
$^{172}_{\,\,\,79}$Au$^{*}$& 5 &1.170  & \cite{Audi}    &$>$550(50) ms      &139 s      &48 s       &-0.260& 2.144& 1.685\\
$^{173}_{\,\,\,79}$Au      & 0 &0.992  & \cite{173Au}   &$\geq$26.3(12) ms  &7.8 s      &5.1 s      &-1.580& 0.893& 0.710\\
$^{173}_{\,\,\,79}$Au$^{*}$& 5 &1.206  & \cite{173Au}   &$\geq$12.2(1) ms   &50 s       &19 s       &-1.914& 1.699& 1.269\\
\hline
\end{tabular}}
\end{center}
\end{table*}

The good agreement with experimental data of the results provided by the present analytical model encouraged us to make predictions for the half-lives of observed proton emitters with incomplete decay information. This refers to observations where the branching ratio for the proton emission in respect to other decay channels is not known or proton emitters with uncertain angular momentum assignment for the ground state which determines the orbital momentum of the emitted proton. In the first case, we have just a lower bound for the decay half-lives. The theoretical predictions shown in Table \ref{tab:2} for $^{144}$Tm, $^{172}$Au, $^{172}$Au$^{*}$, $^{173}$Au, and $^{173}$Au$^{*}$ nuclei are within the corresponding restrictions. Relatively long half-lives are obtained for both ground state and isomeric proton emissions of the $^{172}$Au isotope, which is partly due to unusually low reported $Q_{p}$ values. For the proton emissions with uncertain orbital momentum, we provided predictions in Table \ref{tab:2} for most probable $l$ values. In this way one can ascertain the most likely angular momentum state of the ground state for the nuclei under consideration. The ground state for the two lighter nuclei $^{117}$La and $^{121}$Pr is predominantly considered to be $3/2^{+}$, which corresponds to $l=2$. This assignment is based on the theoretical reproduction of the experimental half-live, which is however model dependent. More recent theoretical calculations based on the consistent treatment of Coriolis and pairing interactions \cite{117La2,121Pr2} point to a $7/2^{-}$ ground state associated to an $l=3$ emitted proton for these two nuclei. This choice is also suggested by the present model predictions and even stronger by the UDLP calculations. This is a good example for how the intrinsic simplicity of the proton emission in comparison to preformed clusters of nucleons can be exploited to obtain unique spectroscopic information on the quantum states of nuclei.

\begin{figure*}[t!]
\begin{center}
\includegraphics[clip,trim = 30mm 6mm 30mm 20mm,width=0.48\textwidth]{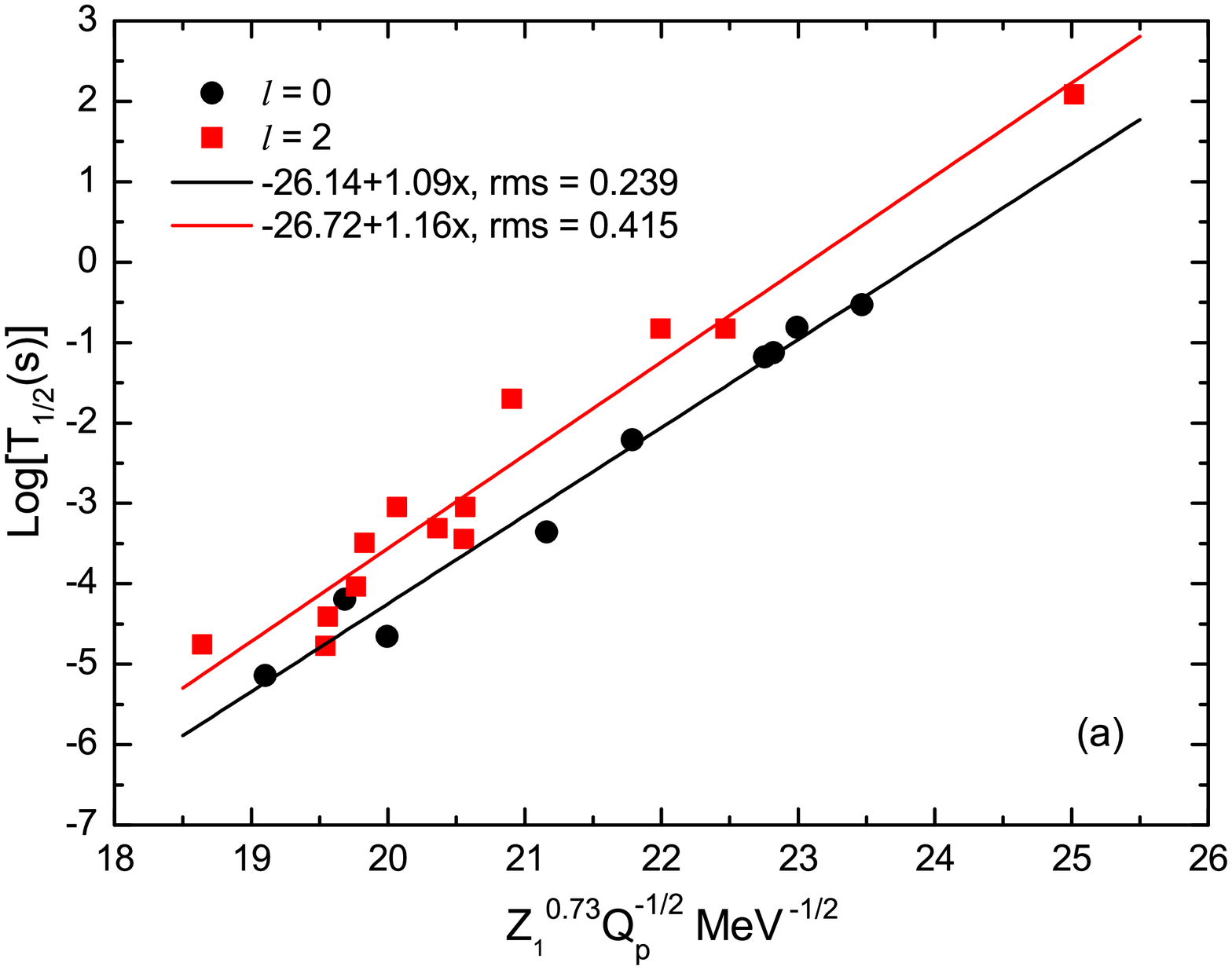}
\includegraphics[clip,trim = 30mm 6mm 30mm 20mm,width=0.48\textwidth]{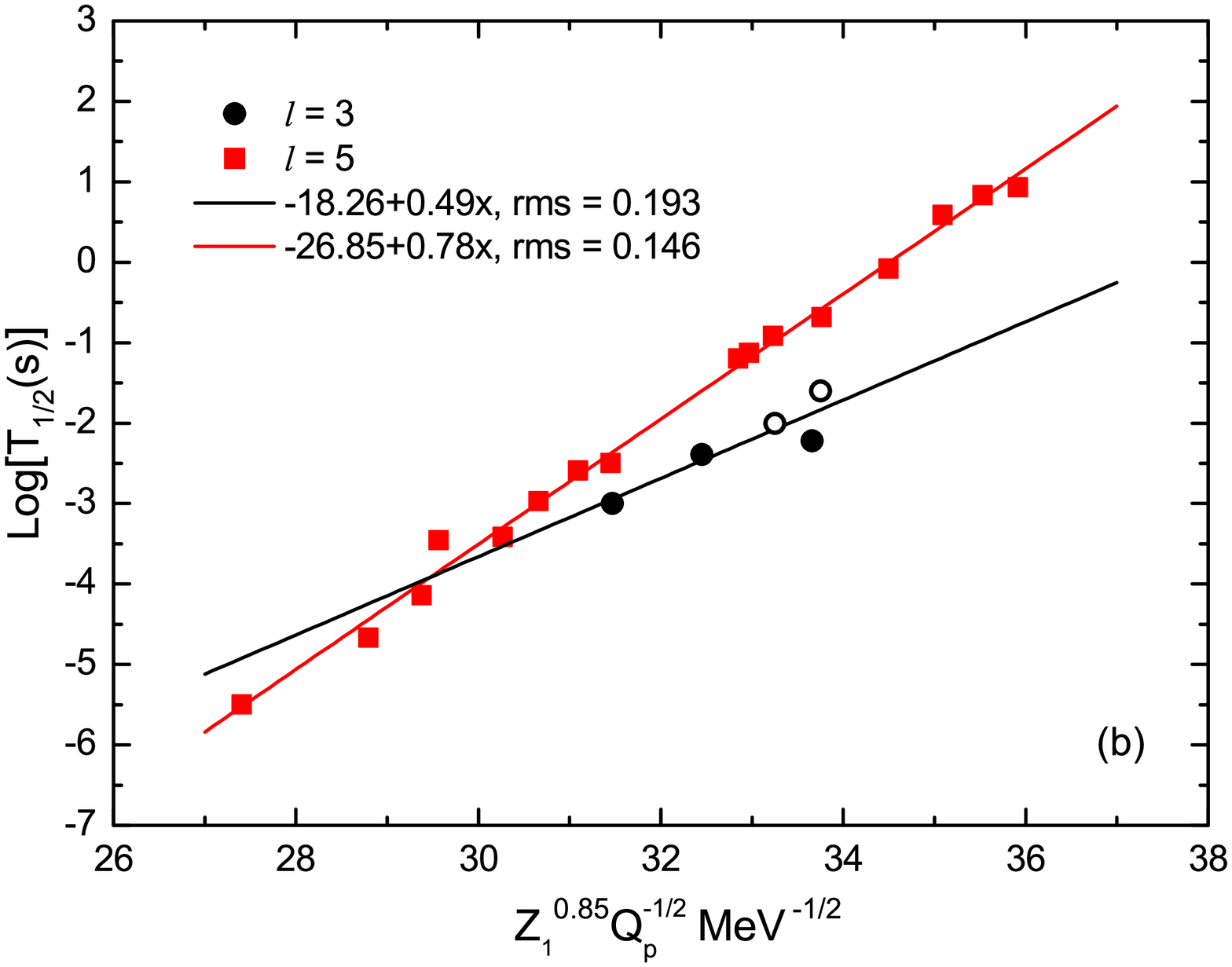}
\caption{Experimental half-lives for proton emission plotted as a function of $Z_{1}^{0.73}/\sqrt{Q_{p}}$ for even $l$ (a), and as a function of $Z_{1}^{0.85}/\sqrt{Q_{p}}$ for odd $l$ data points (b). $Z_{1}$ is the charge number of the daughter nucleus. Data points corresponding to each value of orbital momentum $l$ are denoted by different symbols. The straight lines represent linear fits. The two open circles in (b) denote the values for $^{117}$La and $^{121}$Pr nuclei, added after analysing their theoretical predictions.}
\label{lbrown}
\end{center}
\end{figure*}

An alternative means to confirm specific ingredients of the proton decay is given by empirical correlations. There are few such formulations for the proton decay where the orbital momentum dependence is considered scalable \cite{Medeiros,Qi,Qian,Ni,Delion2}. These correlations are quite accurate in what concerns the systematization of experimental data by orbital angular momentum for two specific ranges of charge number, $Z<68$ and $Z>68$ \cite{Delion1,Delion2} which exhibit distinct deformation properties. Indeed, the lighter emitting nuclei have pronounced prolate deformation, while the heavier ones are predominantly spherical with oblate tendencies. Here we will employ a Brown-type empirical formula \cite{Brown,Horoi,And1,And2,Andreea3}:
\begin{equation}
\log T_{1/2}(s)=\frac{\alpha Z_{1}^{\beta}}{\sqrt{Q_{p}}}+\gamma,
\label{brown}
\end{equation}
to obtain correlations specific to the most common orbital momentum values found in the whole set of proton emitters regardless of their charge number and consequently deformation. $\alpha,\beta$ and $\gamma$ are fitting parameters. In a representation in terms of the quantity $Z_{1}^{\beta}/\sqrt{Q_{p}}$, the above formula is just a straight line with a slope $\alpha$ and intercept $\gamma$. When $\beta=0$, one recovers the well known Geiger-Nuttall law \cite{Geiger}. While the value $\beta=1$, transforms the first term of Eq.(\ref{brown}) into just the Coulomb parameter $Z_{1}/\sqrt{Q_{p}}$ for a proton-nucleus system. The Coulomb parameter is a very often-used variable for various decay laws \cite{Wang2015} and is a part of the universal decay law valid for all kind of clusters and for all isotopic series \cite{UDL1,UDL2} as well as its generalization to the proton emission \cite{Qi}. Therefore, an intermediate $\beta$ value serves as a natural interpolation \cite{Brown} between equally successful linear plots of the Geiger-Nuttall and universal decay laws. Numerical applications \cite{Brown,Horoi,Andreea3} showed that the optimal value of $\beta$ resides in the vicinity of 0.5 for cluster emissions. Contrary to the cluster radioactivity where the centrifugal contribution can be justifiably neglected, the proton emission is very sensitive to the value of the orbital angular momentum. Nevertheless, fitting the experimental proton emission data corresponding in part to $l=0,2$ and 5 with the formula (\ref{brown}), one observed that the power parameter $\beta$ is essentially the same for even $l$ fits. Moreover, as can be seen from Fig.\ref{lbrown}(a), the fitting lines are almost parallel, with $l=2$ line positioned above the $l=0$ one. Within such a systematics, the $l=0$ and $l=2$ data sets are quite distinguishable, with a slight superposition of data points corresponding to highest $Q_{p}$ values. Therefore, Eq.(\ref{brown}) can be used as a reliable test for angular momentum assignment to proton emitter states. In what concerns the odd $l$ nuclei, the fit on $l=5$ data revealed a higher value for the power parameter which provided in Fig.\ref{lbrown}(b) an impressive linear distribution of the corresponding data points. On the other hand, the fitting of the few $l=3$ half-live values is far from being concluding. Nevertheless, including the $l=3$ predictions for $^{117}$La and $^{121}$Pr from Table \ref{tab:2} and fitting the data against the same abscissa, as in the $l=5$ case, one obtained a reasonable linear dependence. This result, once again, supports the $l=3$ proton emission for $^{117}$La and $^{121}$Pr nuclei, whose data points would be otherwise completely out of the linear correlation of $l=2$ from Fig.\ref{lbrown}(a). The difference between the odd $l$ slopes generates also a possible superposition of $l=3$ and $l=5$ results for high $Q_{p}$ values just like in the even $l$ case of Fig.\ref{lbrown}(a). The similarity with the even $l$ case is also reflected in the relative position of lines, {\it i.e.} the higher $l$ line is generally above.

\section{Conclusions}

In summary, we constructed a simple analytical model for the proton decay based on the WKB approximation. The WKB formula was used to calculate the penetrability of a phenomenological barrier mainly defined by the centrifugal and electrostatic contributions. The barrier is considered with a pre-scission part defining the probability for a proton to reach the touching configuration. The novelty of the present approach consists in the generalization of the usual Coulomb electrostatic interaction by means of the Hulthen potential which has a shorter range specified by its screening parameter $a$. Such a potential is specifically suited for the proton emission where due to low reaction energies, the tail of the potential barrier acquires a significant role. The simple structure of the proposed model provides an analytical formula for the proton emission half-time depending only on $a$ and other decay information. The screening parameter $a$ is fixed by fitting 41 experimental data points. The agreement with experimental data is fairly good considering that we have a single adjustable parameter. In this sense, the theoretical predictions were also confronted with the results of the universal decay law for the proton emission which exhibits a similar agreement with experiment but is employing four adjustable parameters. This speaks for the ability of the screening parameter to account for the missing secondary ingredients which might affect the proton emission. The model is used to make some predictions, which proved especially useful in assigning the proton orbital momentum in case of the $^{117}$La and $^{121}$Pr decaying nuclei. An important byproduct of this study is the proposal of a new empirical correlation between the half-lives for the proton emission, the charge number of the daughter nucleus and the $Q_{p}$ value, which is differentiated by the proton orbital momentum. The last aspect can be used as a reliable tool to assign the angular momentum and parity of the proton decaying states.
\\

\acknowledgments
The authors acknowledge the financial support received from the Romanian Ministry of Education and Research, through the Project PN-16-42-01-01/2016.

\end{document}